\documentstyle[12pt,axodraw,epsfig]{article}
\def\be{\begin{equation}}
\def\ee{\end{equation}}
\def\bea{\begin{eqnarray}}
\def\eea{\end{eqnarray}}

\def \lsim{\mathrel{\vcenter
     {\hbox{$<$}\nointerlineskip\hbox{$\sim$}}}}

\begin{document}

\begin{flushright}
TIFR/TH/02-13
\end{flushright}

\vspace*{1cm}

\begin{center}
{\Large\bf Using Tau Polarization as a Distinctive SUGRA Signature at
LHC} \\[1cm]
{\large
Monoranjan Guchait$^a$ and D.P. Roy$^b$ 
}
\vspace*{1cm}

{\it
\noindent $^a$Department of High Energy Physics\\
\vspace*{0.2cm}
\noindent $^b$Department of Theoretical Physics\\
\vspace{0.2cm}
Tata Institute of
Fundamental Research, Homi Bhabha Road\\
Mumbai 400 005, INDIA}
\end{center}
\vspace*{1cm}

\begin{center}
\underbar{\bf Abstract}
\end{center}
\bigskip

In the minimal SUGRA model the lighter tau slepton is expected to be
the second lightest superparticle over a large parameter range at
large $\tan\beta$.  Consequently one expects a viable SUGRA signal at
LHC in the tau lepton channel coming from the decay of these tau
sleptons.  The model predicts the polarization of this tau lepton to
be +1 to a very good accuracy.  We show how this prediction can be
tested by looking at the momentum fraction of the tau-jet, carried by
the charged prong, in its 1-prong hadronic decay channel.

\newpage 

The minimal supergravity (SUGRA) model provides a well-motivated and
economical parametrisation of the minimal supersymmetric standard
model (MSSM) in terms of the four continuous and one discrete
parameters [1] 
\be
m_{1/2},m_0,A_0,\tan\beta \ {\rm and} \ {\rm sgn} (\mu).
\label{one}
\ee
The mass parameter $\mu$ represents the mixing between the two Higgs
doublets, while $\tan\beta$ denotes the ratio of their vacuum
expectation values.  The magnitude of $\mu$ is determined by the
electroweak symmetry breaking condition.  The other mass parameters
$m_{1/2},m_0$ and $A_0$ denote the common gaugino and scalar masses
and the common trilinear coupling term at the unification scale.  In
fact $m_{1/2}$ continues to represent the rough magnitude of the
$SU(2)$ gaugino mass $M_2$ down to the weak scale, i.e.
\be
m_{\tilde W_1},m_{\tilde Z_2} \simeq M_2 \simeq 0.8 m_{1/2},
\label{two}
\ee
where $\tilde W_1$ denotes the lighter chargino and $\tilde Z_2$ the
second lightest neutralino.  Similarly $m_0$ continues to represent
the rough magnitude of the right-handed slepton masses down to the weak scale, i.e. 
\be
m^2_{\tilde \ell_R} \simeq m^2_0 + 0.15 m^2_{1/2},
\label{three}
\ee
neglecting the Yukawa coupling contribution to the RGE.  However at
large $\tan\beta$ the Yukawa coupling contribution drives the
$\tilde\tau_R$ mass significantly below (\ref{three}).  Moreover there
is a significant mixing between the $\tilde\tau_{L,R}$ states, as
represented by the off-diagonal term
\be
m^2_{LR} = - m_\tau (A_\tau + \mu \tan \beta),
\label{four}
\ee
which drives the lighter mass eigen value further down.  Thus the
lighter stau mass eigen state 
\be
\tilde\tau_1 = \tilde\tau_R \sin\theta_\tau + \tilde\tau_L \cos\theta_\tau
\label{five}
\ee
is predicted to be significantly lighter than the other sleptons at
large $\tan\beta$.  Moreover one sees from
eqs. (\ref{two},\ref{three}) that for $m_0 \leq m_{1/2}$ it is expected
to be lighter than $\tilde W_1$ and $\tilde Z_2$ as well, which makes
it the second lightest superparticle after $\tilde Z_1$ [2].  The
cascade decay of superparticles via $\tilde \tau_1$ leads to
\be
\tilde\tau_1 \rightarrow \tau \tilde Z_1.
\label{six}
\ee
Thus the channels containing $\tau$ lepton(s) plus missing $E_T$
$({E\!\!\!/}_T)$ have attracted a great deal of attention as promising
channels for SUGRA signal, particularly for large values of
$\tan\beta$ [2-6].

We consider here a distinctive SUGRA prediction for the polarization
of the $\tau$ resulting from $\tilde\tau_1$ decay (\ref{six}), i.e.
\be
P_\tau = + 1,
\label{seven}
\ee
which has not received much attention so far [7].  As we shall see
below, this is a robust prediction of the minimal SUGRA model, which
holds to a very good precision over the canonical ranges of the model
parameters (\ref{one}).  Of course its practical utility as a
distinctive SUSY signal is restricted to those parameter ranges where
the superparticle decays have a large branching fraction into the
$\tau$ channel (\ref{six}), i.e. large $\tan\beta$ and $m_0 \leq
m_{1/2}$.  In this case the identification of $\tau$ in its hadronic
decay channel at LHC will enable us to measure its polarization and
test the above prediction rather unambiguously.  Note that in contrast
to the SUGRA prediction of $P_\tau = +1$, the SM background from $W
\rightarrow \tau\nu$ and $Z (H^0) \rightarrow \tau^+ \tau^-$ decays
predict $P_\tau = -1$ and $0$ respectively.  Of course there is an
alternative source of $P_\tau = +1$ in the MSSM, i.e. the charged
Higgs decay $H^\pm \rightarrow \tau^\pm \nu$ [8,9].  However it will
not pose a serious problem for the SUGRA signal at least in the
$\tau\tau$ channel considered below, since the pair production of
charged Higgs boson has a negligible cross-section at LHC.  Thus the
experimental confirmation of the predicted $\tau$ polarization will
provide a distinctive test for the minimal SUGRA model.

The polarization formalism for $\tau$ leptons in SUSY cascade decay
has been discussed in ref. [10].  The polarization of the $\tau$ for
$\tilde\tau_1$ decay (\ref{six}) is given in the collinear
approximation $(m_\tau \ll m_{\tilde\tau_1})$ by
\bea
P_\tau &=& {(a^R_{11})^2 - (a^L_{11})^2 \over (a^R_{11})^2 +
(a^L_{11})^2}, \nonumber \\[2mm] 
a^R_{11} &=& -{2g \over \sqrt{2}} N_{11} \tan\theta_W \sin\theta_\tau
- {g m_\tau \over \sqrt{2} m_W \cos\beta} N_{13} \cos\theta_\tau,
\nonumber \\[2mm] 
a^L_{11} &=& {g \over \sqrt{2}} \left[N_{12} + N_{11}
\tan\theta_W\right] \cos\theta_\tau - {g m_\tau \over \sqrt{2} m_W
\cos\beta} N_{13} \sin\theta_\tau, 
\label{eight}
\eea
where
\be
\tilde Z_1 = N_{11} \tilde B + N_{12} \tilde W + N_{13} \tilde H_1 +
N_{14} \tilde H_2.
\label{nine}
\ee
At low $\tan\beta$ the stau mixing angle $\cos\theta_\tau$ is small,
so that the first term of $a^R_{11}$ dominate over the others.
Consequently $P_\tau = +1$ holds to very good precision.  But the SUSY
cascade decay into the $\tau$ channel (\ref{six}) becomes large only
for a small part of the parameter space $(m_0 \sim {1\over2} m_{1/2})$
at low $\tan\beta$.  In the high $\tan\beta$ region of our interest
both the stau mixing angle $\cos\theta_\tau$ and the higgsino coupling
factor in eq. (\ref{eight}) become significant.  However there is an
effective cancellation between the two terms in $a^L_{11}$, which
ensures $a^L_{11} \ll a^R_{11}$.  Consequently $P_\tau = +1$ holds to a very
good approximation in the high $\tan\beta$ region as well.

In Figs. 1 and 2 we have shown the regions of large $P_\tau$ and large
$\tilde W_1 \rightarrow \tilde\tau_1 \nu$ branching fraction for
different sets of SUGRA parameters using the ISASUGRA code-version
7.48 [11].  Fig. 1 shows that at $\tan\beta = 30$ we have $P_\tau > 0.9$ over
almost the entire $m_0 - m_{1/2}$ plane for both signs of $\mu$.
Fig. 2 shows that this remains true when we change the trilinear
coupling parameter from 0 to 500 GeV or the $\tan\beta$ value from 30
to 40.  Thus the $\tau$ polarization prediction (\ref{seven}) for the
decay channel (\ref{six}) is seen to be a very robust prediction of
the minimal SUGRA model.

Figs. 1 and 2 show the contours for large branching fractions of 
\be
\tilde W_1 \rightarrow \tilde\tau_1 \nu,
\label{ten}
\ee
which is the most important source of $\tau$ production in the SUGRA
model.  They also show the threshold contour for this two-body decay
channel along with that of the competing channel, $\tilde W_1
\rightarrow W\tilde Z_1$.  The region, $m_0 \lsim m_{1/2}$,
corresponds to large branching fractions for the decay process
(\ref{ten}) and hence to a large SUGRA signal in the $\tau$ channel.
We have checked that the $\tau$ polarization remains $> 0.95$ over
this region.

For a detailed investigation of the SUGRA signal in the $\tau\tau$
channel and the effect of $\tau$ polarization we have selected a
representative point from Fig. 2b, i.e.
\be
m_0 = 250 \ {\rm GeV}, m_{1/2} = 300 \ {\rm GeV}, A_0 = 0, \tan\beta =
40, {\rm Sgn} (\mu) = +{\rm ve}.
\label{eleven}
\ee
The resulting superparticle masses (in GeV) and mixing angle are
\be
m_{\tilde g} = 736, m_{\tilde q} = 680, m_{\tilde W} = 232, m_{\tilde
Z_2} = 233, m_{\tilde\tau_1} = 197, m_{\tilde Z_1} = 124,
\cos\theta_\tau = 0.41.
\label{twelve}
\ee
The relevant decay branching fractions are
\bea
B(\tilde q \rightarrow q \tilde W_1, q\tilde Z_2,q\tilde Z_1) &=&
0.62,0.30,0.08 \nonumber \\[2mm] 
B(\tilde W_1 \rightarrow \tilde\tau_1 \nu \rightarrow \tau \nu\tilde
Z_1) &=& 0.76, B(\tilde Z_2 \rightarrow \tau' \tilde\tau_1 \rightarrow
\tau' \tau \tilde Z_1) = 0.97.
\label{thirteen}
\eea
The two main sources of $\tau\tau$ signal are
\be
\tilde q \bar{\tilde q} \rightarrow q \bar q \tilde W_1^+ \tilde W^-_1
\rightarrow q\bar q \nu\bar\nu \tilde Z_1\tilde Z_1 \tau\tau,
\label{fourteen}
\ee
\be
\tilde q \bar{\tilde q} \rightarrow q \bar q \tilde W_1^\pm \tilde Z_2
\rightarrow q\bar q \tilde Z_1 \tilde Z_1 \tau' \tau \tau,
\label{fifteen}
\ee
which have branching fractions of 22 and 27\% respectively.  It may be
noted here that $P_\tau = 0.985$, but $P_{\tau'} = -0.996$.  The
dominant contribution to the $\tau'$ polarization comes from $a^L_{12}
\simeq {g \over \sqrt{2}} N_{22} \cos\theta_\tau$ [10]. However we shall see
below that the presence of $\tau'$ shall not seriously compromise the
experimental test of $P_\tau$.

We have done a parton level Monte Carlo study of the SUGRA signal from
$\tilde g\tilde g$ and $\tilde g\tilde q$ production in the channels
(\ref{fourteen}) and (\ref{fifteen}).  In addition to the taus and the
${E\!\!\!/}_T$ there are at least three hadronic jets.  To simulate
detector resolution we have applied Gaussian smearing on each jet
$p_T$ (including the $\tau$-jets) with 
\be
(\sigma(p_T)/p_T)^2  = (0.6/\sqrt{p_T})^2 + (0.04)^2
\label{sixteen}
\ee
in GeV units.  The ${E\!\!\!/}_T$ is evaluated from the vector sum of
the jet $p_T$'s after resolution smearing. To suppress the SM
background we require the identified $\tau$-jet pair to be accompanied
by at least three hard jets and a large ${E\!\!\!/}_T$ with 
\be
{E\!\!\!/}_T,p_{T_1} > 100 \ {\rm GeV}, p_{T_2},p_{T_3} > 50 \ {\rm
GeV}, M_{eff} > 500 \ {\rm GeV},
\label{seventeen}
\ee
where $M_{eff}$ is the scalar sum of all the jet $p_T$'s and
${E\!\!\!/}_T$ [6].

The hadronic decay channel of $\tau$ is known to be sensitive to
$\tau$ polarization [8,9].  We shall concentrate on the 1-prong
hadronic decay of $\tau$, which is best suited for $\tau$
identification.  It accounts for 80\% of hadronic $\tau$ decay and
50\% of its total decay width.  The main contributors to the 1-prong
hadronic decay are
\be
\tau^\pm \rightarrow \pi^\pm \nu (12.5\%), \rho^\pm \nu (26\%),
a^\pm_1 \nu (7.5\%),
\label{eighteen}
\ee
where the branching fractions for $\pi$ and $\rho$ include the small
$K$ and $K^\star$ contributions respectively, which have identical
polarization effects [12].  Together they account for 90\% of the
1-prong hadronic decay.  The CM angular distribution of $\tau$ decay
into $\pi$ or a vector meson $v (= \rho,a_1)$ is simply given in terms
of its polarization as
\bea
{1 \over \Gamma_\pi} {d\Gamma_\pi \over d\cos\theta} &=& {1\over2} (1
+ P_\tau \cos\theta),
\label{ninteen} \\[2mm]
{1 \over \Gamma_v} {d\Gamma_{v \ L,T} \over d\cos\theta} &=&
{{1\over2} m^2_\tau,m^2_v \over m^2_\tau + 2m^2_v} (1 \pm P_\tau
\cos\theta), 
\label{twenty}
\eea
where $L,T$ denote the longitudinal and transverse polarization states
of the vector meson.  The fraction $x$ of the $\tau$ lab. momentum
carried by its decay meson is related to the angle $\theta$ via
\be
x = {1\over 2} (1 + \cos\theta) + {m^2_{\pi,v} \over 2m^2_\tau} (1 -
\cos\theta) 
\label{twentyone}
\ee
in the collinear approximation.  The only measurable $\tau$ momentum
is the visible momentum of the $\tau$-jet,
\be
p_{\tau{\rm -jet}} = x p_\tau.
\label{twentytwo}
\ee
We see from eqs. (\ref{ninteen}-\ref{twentytwo}) that $P_\tau = +1$
gives a harder $\tau$-jet than $P_\tau = -1$.

We shall require two identified $\tau$-jets with
\be
p^T_{\tau{\rm -jet}} > 40 \ {\rm GeV}, \ |\eta_{\tau {\rm -jet}}| <
2.5, \ R > 0.2,
\label{twentythree}
\ee
where
\be
R = p_{\pi^\pm}/p_{\tau {\rm -jet}},
\label{twentyfour}
\ee
i.e. the fraction of the visible $\tau$-jet momentum carried by the
charged prong.  This can be obtained by combining the charged prong
momentum measurement in the tracker with the calorimetric energy
deposit of the $\tau$-jet.  This quantity $R$ is a good discriminator
of $\tau$ polarization [9].

Before proceeding to the test of $\tau$-polarization, however, let us
discuss the size of the $\tau\tau$ signal and also the contamination
of $P_\tau$ by the presence of $\tau'$.  With the kinematic cuts of
(\ref{twentythree}) the CDF colaboration has estimated a $\tau$
detection efficiency of 50\% and a fake tau rejection factor of 0.1\%
at Tevatron [13]; and one expects similar numbers for the CMS
experiment at LHC [14].  With the SM background already suppressed by
the ${E\!\!\!/}_T$ and $M_{\rm eff}$ cuts of (\ref{seventeen}) it
should be possible to get a better efficiency for $\tau$ detection at
the cost of a lesser rejection factor for fakes.  However we shall
conservatively assume a 50\% detection efficiency for each
$\tau$-jet.  Combining this with the 50\% branching fraction for
1-prong hadronic $\tau$ decay and the 22 and 27\% branching fractions
for the decay chains (\ref{fourteen}) and (\ref{fifteen}) gives a net
probability factor of about 1.5\% for catching the SUGRA signal in the
two indentified $\tau$-jets channel via each of these decay chains.
The effect of the kinematic cuts (\ref{seventeen}) and
(\ref{twentythree}) are taken into account through the Montecarlo
program.  We get a net signal cross-section of $\sim 15$ fb in this
channel from $\tilde g\tilde g$ and $\tilde q\tilde g$ production,
which includes an average $K$ factor of 1.4 from NLO correction [15].
It corresponds to 150 events for the low luminosity $(10 \ {\rm
fb}^{-1}/{\rm yr})$ run of LHC, going up to 1500 events at high
luminosity $(100 \ {\rm fb}^{-1}/{\rm yr})$.  

We have tried to minimise the contamination of the $\tau$ polarization
by the presence of the $\tau'$ in the decay chain (\ref{fifteen}) via
the following requirements.  Firstly the two indentified $\tau$-jets
are required to have an invariant mass $> 100$ GeV, which ensures that
they come from different gauginos.  Secondly they are required to have
opposite charge.  This ensures that the detection probability for the
$\tau\tau$ and $\tau\tau'$ pair in (\ref{fifteen}) is 1/2 each, while
it is 1 for the $\tau\tau$ pair in (\ref{fourteen}).  Moreover the
probability of the $\tau$-jet passing the visible $p_T > 40$ GeV cut
(\ref{twentythree}) is only 25\% for the $\tau'$ against 60\% for
$\tau$.  The difference comes partly due to their opposite
polarization and partly due to the different kinematics.  The end
result is that the relative probability factors of detecting the
$\tau\tau$ and $\tau\tau'$ pairs from (\ref{fourteen},\ref{fifteen})
are $.60 \times (.22 + .27/2)$ and $.25 \times (.27/2)$ respectively.  
One can
easily check that it has the effect of reducing the predicted
polarization (\ref{seven}) by 13-14\% for each $\tau$.  We shall
neglect this effect as a first approximation.  Note that the
requirement of opposite charge for the pair of $\tau$-jets also cuts
the fake $\tau$ background by half.

To test $\tau$-polarization we observe from
eqs. (\ref{ninteen}-\ref{twentytwo}) that the hard $\tau$-jet is
dominated by the $\pi$ and longitudinal vector meson $(\rho_L,a_{1L})$
contributions for $P_\tau = +1$, while it is dominated by the
transverse $\rho$ and $a_1$ contributions for $P_\tau = -1$.  The two
sets can be distinguished by exploiting the fact that the transverse
$\rho$ and $a_1$ decays favour even sharing of momentum among the
decay pions, while the longitudinal $\rho$ and $a_1$ decays favour
uneven sharing, where the charged pion carries either very little or
most of the momentum.  It is easy to derive this quantitatively for
$\rho$ decay; but one has to assume a dynamical model for $a_1$ decay
for a quantitative result.  We shall assume the model of ref. [16],
based on conserved axial-vector current approximation, which provides
an accurate description of the $a_1 \rightarrow 3\pi$ decay data.  A
detailed account of the $\rho$ and $a_1$ decay formalisms including
finite width effects can be found in [8,9].  A simple FORTRAN code for
1-prong hadronic decay of polarized $\tau$ based on these formalisms
can be obtained from one of the authors (D.P. Roy).

Figs. 3 shows the $P_\tau = +1$ signal as a function of the $\tau$-jet
momentum fraction $R$, carried by the charged-prong.  For comparison
it also shows the corresponding distribution assuming the signal to
have $P_\tau = -1$.  This could be the case e.g. in some nonuniversal
SUSY models with a higgsino LSP or the anomaly mediated SUSY model
with a wino LSP. In particular the decay of a stau into tau and a wino 
LSP will always proceed via its left component giving $P_\tau$=-1.
The $P_\tau = +1$ signal shows the peaks at the two
ends from the $\rho_L,a_{1L}$ along with the pion contribution (added
to the last bin).  Of course the peak at the low $R$ end is partly cut
out by the $\tau$-identification requirement (\ref{twentythree}).  In
contrast the $P_\tau = -1$ distribution shows a central peak due to
the $\rho_T,a_{1T}$ along with a reduced pion contribution.  The
complimentary shape of the two distributions reflects the uneven
sharing of momentum between the charged and the neutral pions for
$\rho_L,a_{1L}$ decays and its even sharing for $\rho_T,a_{1T}$
decays.  The distribution for $P_\tau = 0$, lying midway between the
above two distributions, is also shown for comparison.

Let us look at the fractional cross-section lying above $R = 0.8$ as a
measure of the $\tau$ polarization.  This fraction is 0.60 for $P_\tau
= +1$ but only .25 (.40) for $P_\tau = -1 (0)$.  Thus one can
require both the identified $\tau$-jets to contain hard charged prongs
carrying $> 80\%$ of the respective $\tau$-jet momenta, i.e. $R_{1,2}
> 0.8$.  Then 36\% of the $P_\tau = +1$ signal will pass this cut,
while the corresponding fraction is only 6 (16)\% for $P_\tau = -1
(0)$.  Thus with the expected sample of $\sim 150$ signal events at
the low luminosity run of LHC $(10 \ {\rm fb}/{\rm yr})$, one expects
54 events to pass this cut for $P_\tau = +1$ against only
9 (24) events for $P_\tau = -1 (0)$.  This will provide an
unambiguous test for the $\tau$-polarization, predicted by the minimal
SUGRA model.  At the high luminosity run the test can be extended to a
wider range of the parameter space of figs. 1 and 2, going down to $B$
$(\tilde W_1 \rightarrow \tilde\tau_1 \nu \rightarrow \tau\nu
\tilde Z_1) = 0.3$ and upto $m_{1/2} = 400 - 500$ GeV.  One can extend
this range further in the one identified $\tau$ channel.  It may be
added here that for the fake $\tau$ background from QCD jets the
fractional cross-section surviving the $R > 0.8$ cut is even less than
0.2 [14].  Thus the $R > 0.8$ peak provides a distinctive test for
the $P_\tau = +1$ signal not only against the $P_\tau = -1 (0)$
background but against the fake $\tau$ background as well. Finally,
it should be noted that while we have focussed the curreny analysis 
on the SUGRA model the same polarization strategy can be used to 
distinguish the SUSY signal from the SM backgrounds in the gauge
mediated SUSY breaking model[17], where one expects $P_\tau$=+1
$\tilde\tau_R \rightarrow \tau \tilde G$ decay.

We thank Manuel Drees and Kaoru Hagiwara for discussions on different
aspects of tau polarization.


\section*{\bf References}
\medskip

\begin{enumerate}
\item[{[1]}] For reviews see H.P. Nilles, Phys. Rep. 110 (1984) 1;
H.E. Haber and G.L. Kane, Phys. Rep. 117 (1985) 75.
\item[{[2]}] M. Drees and M.M. Nojiri, Nucl. Phys. B369 (1992) 54.
\item[{[3]}] H. Baer, C.H. Chen, M. Drees, F. Paige and X. Tata,
Phys. Rev. D58 (1998) 075008.
\item[{[4]}] K.T. Matchev and D.M. Pierce, Phys. Rev. D60 (1999)
075004; J.D. Lykken and K.T. Matchev, Phys. Rev. D61 (2000). 
\item[{[5]}] V. Barger, C. Kao and T. Li, Phys. Lett. B433 (1998) 328;
V. Barger and C. Kao, Phys. Rev. D60 (1999) 115015.
\item[{[6]}] H. Baer, C.H. Chen, M. Drees, F. Paige and X. Tata,
Phys. Rev. D59 (1999) 055014; I. Hinchliffe and F. Paige,
Phys. Rev. D61 (2000) 095011.
\item[{[7]}] M. Guchait and D.P. Roy, hep-ph/0109096, Phys. Lett. B
(in press).
\item[{[8]}] B.K. Bullock, K. Hagiwara and A.D. Martin,
Phys. Rev. Lett. 67 (1991) 3055; Nucl. Phys. B395 (1993) 499.
\item[{[9]}] S. Raychaudhuri and D.P. Roy, Phys. Rev. D52 (1995) 1556;
D53 (1996) 4902; D.P. Roy, Phys. Lett. B459 (1999) 607.
\item[{[10]}] M.M. Nojiri, Phys. Rev. D51 (1995) 6281; M.M. Nojiri,
K. Fujii and T. Tsukamoto, Phys. Rev. D54 (1996) 6756.
\item[{[11]}] H. Baer, F. Paige, S.D. Protopopescu and X. Tata,
hep-ph/0001086. 
\item[{[12]}] Review of Particle Properties, Euro. Phys. J C15 (2000)
1.
\item[{[13]}] CDF Collaboration: F. Abe et. al. Phys. Rev. Lett. 79
(1997) 3585; M. Hohlmann, preprint Fermilab-conf-98-376-E and
Ph.D. Thesis, Univ. of Chicago (1997).
\item[{[14]}] R. Kinnunen and D. Denegri, CMS Note 1999/037.
\item[{[15]}] W. Beenakker, R. Hopker, M. Spira and P.M. Zerwas,
Nucl. Phys. B492 (1997) 51.
\item[{[16]}] J.H. Kuhn and A. Santamaria, Z. Phys. C48 (1990) 445.
\item[{[17]}] D. A. Dicus, B. Dutta and S. Nandi, Phys. Rev. D56 (1997)
5748; B. Dutta, D. J. Muller and S. Nandi, Nucl. Phys. B544 (1999) 451;
H. Baer et. al., Phys. Rev. D60 (1999) 055001, Phys. Rev. D62 (2000)
095007.
\end{enumerate}

\newpage
\begin{figure}[htbp]
\begin{center}
\vskip-8cm
\hskip-1cm\centerline{\epsfig{file=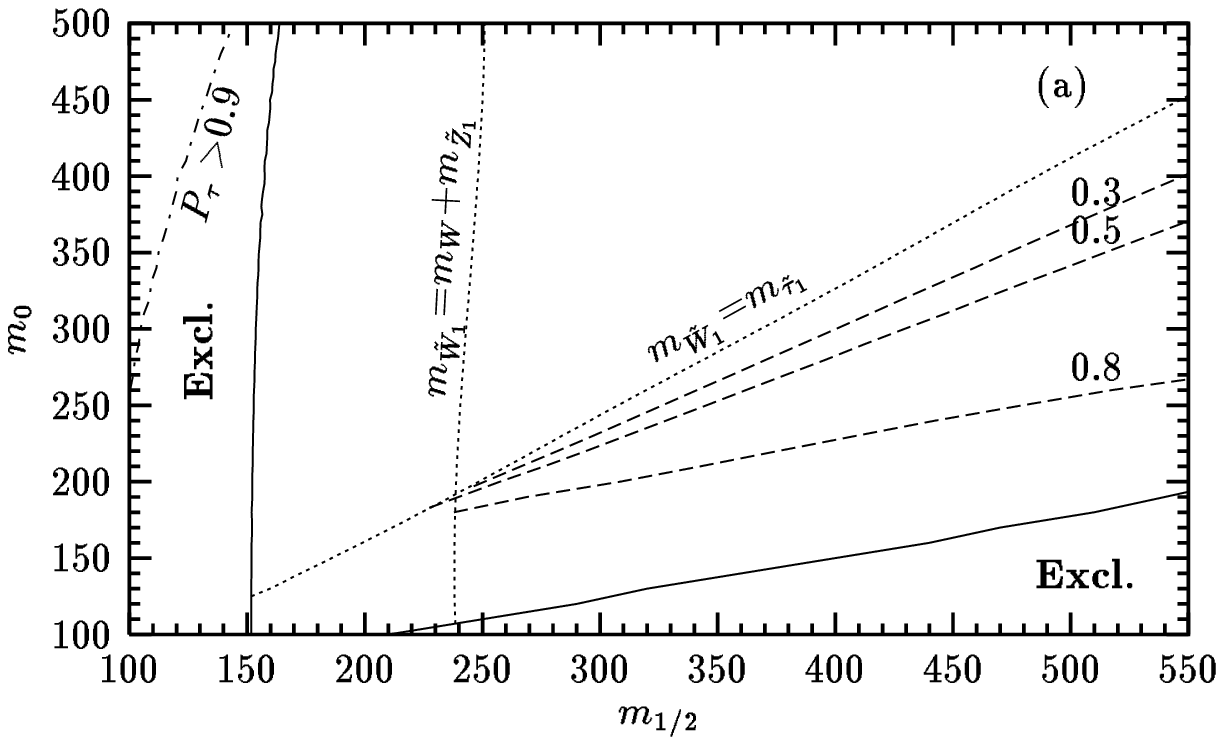,width=23cm}}
\vskip-24cm
\hskip-1cm\centerline{\epsfig{file=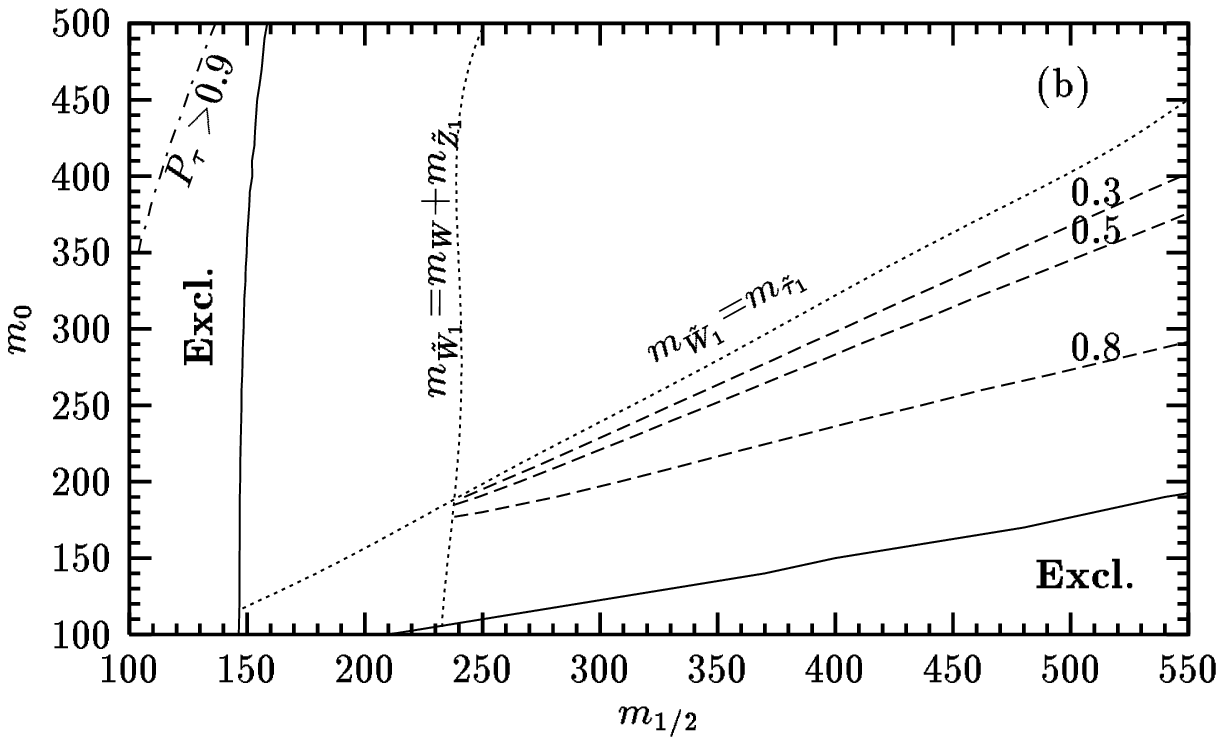,width=23cm}}
\vskip-19cm
\end{center}
\caption{$BR(\tilde W_1 \rightarrow \tilde\tau_1 \nu_\tau)$ is shown
as contour plots(dashed lines) in $m_0$ and $m_{1/2}$ plane for 
$A_0$ = 0, $\tan\beta$ = 30 and 
(a) positive $(\mu)$ (b) negative $(\mu)$. The kinematic
boundaries(dotted lines) are shown for $\tilde W_1 \rightarrow W \tilde
Z_1 $ and $\tilde W_1 \rightarrow \tilde\tau_1 \nu_\tau$ decay. 
The entire region to the right of the boundary(dot-dashed line) corresponds
to $P_\tau >$0.9.
The excluded
region on the right is due to the $\tilde\tau_1$ being the LSP while that on
the left is due to the LEP constraint $m_{\tilde W_1^\pm} >$102 GeV.
}
\end{figure}
\newpage
\begin{figure}[htbp]
\begin{center}
\vskip-8cm
\hskip-1cm\centerline{\epsfig{file=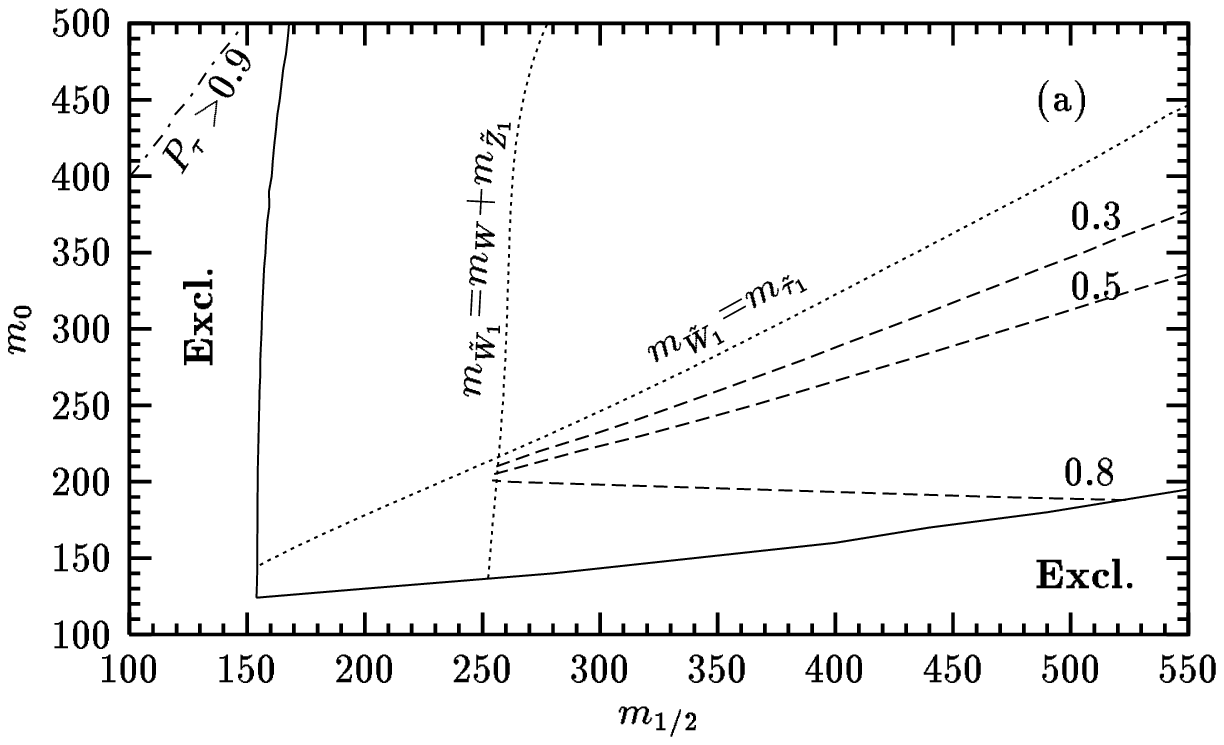,width=23cm}}
\vskip-24cm
\hskip-1cm\centerline{\epsfig{file=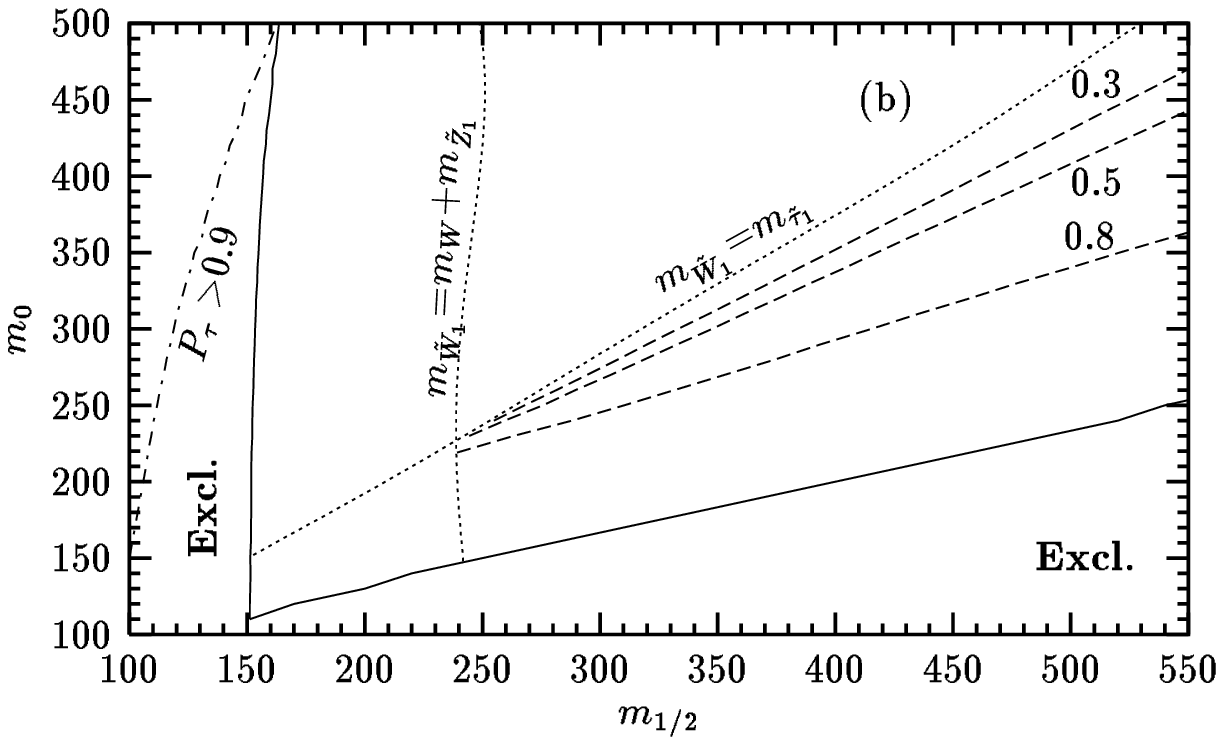,width=23cm}}
\vskip-19cm
\end{center}
\caption{Same as Fig.1 for  
(a) $A_0$ = 500 GeV, $\tan\beta$ = 30 and (b) $A_0$ = 0, $\tan\beta$ = 40 with
positive $(\mu)$ in both cases.
}
\end{figure}
\newpage
\begin{figure}[htbp]
\begin{center}
\vskip-5cm
\hskip-1cm\centerline{\epsfig{file=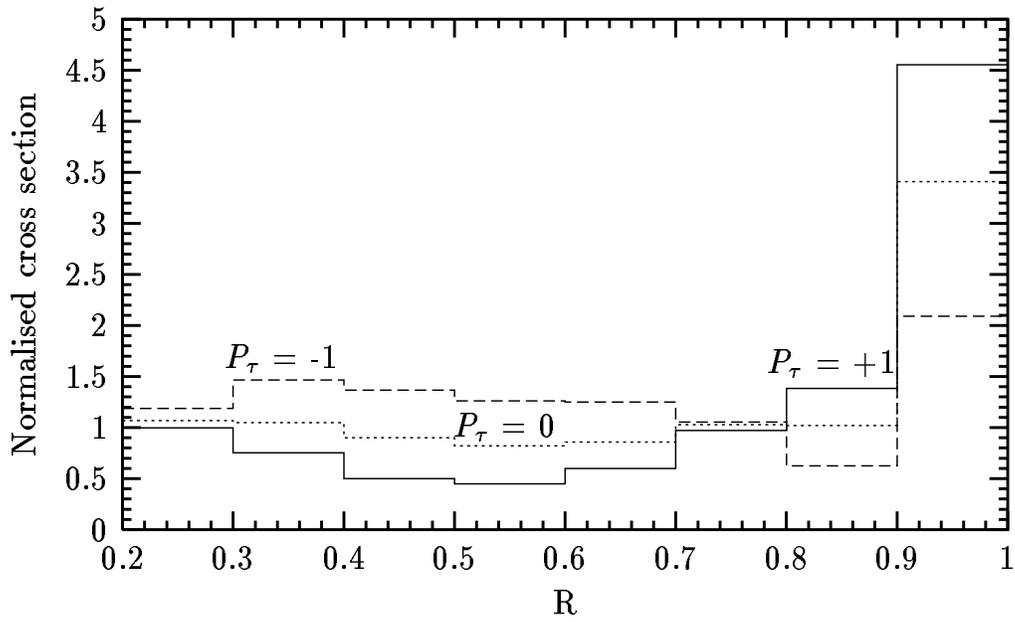,width=23cm}}
\vskip-19cm
\end{center}
\caption{ The normalised SUSY signal cross sections for 
$P_\tau$=1(solid line), 0(dotted
lines) and -1( dashed lines) in the 1-prong hadronic
$\tau$-jet channel shown as functions of the $\tau$ jet momentum fraction 
(R) carried by the charged prong. 
}
\end{figure}

\end{document}